\documentclass[aps,pra,twocolumn,groupedaddress,showpacs]{revtex4-1}
\usepackage{epsfig}
\usepackage{graphics}
\usepackage{amssymb}
\usepackage{amsmath}

\usepackage{epsfig}

\newcommand{\nl}{\nonumber\\}

\newcommand{\ha}{\hat{a}}

\begin{document}

\title{Optimized Heralding Schemes for Single Photons}

\author{Yu-Ping Huang}
\affiliation{Center for Photonic Communication and Computing, EECS Department\\
Northwestern University, 2145 Sheridan Road, Evanston, IL 60208-3118}
\author{Joseph B. Altepeter}
\affiliation{Center for Photonic Communication and Computing, EECS Department\\
Northwestern University, 2145 Sheridan Road, Evanston, IL 60208-3118}
\author{Prem Kumar}
\affiliation{Center for Photonic Communication and Computing, EECS Department\\
Northwestern University, 2145 Sheridan Road, Evanston, IL 60208-3118}

\begin{abstract}
A major obstacle to a practical, heralded source of single photons is the fundamental trade-off between high heralding efficiency and high production rate. To overcome this difficulty, we propose applying sequential spectral and temporal filtering on the signal photons before they are detected for heralding. Based on a multimode theory that takes into account the effect of simultaneous multiple photon-pair emission, we find that these filters can be optimized to yield both a high heralding efficiency and a high production rate. While the optimization conditions vary depending on the underlying photon-pair spectral correlations, all correlation profiles can lead to similarly high performance levels when optimized filters are employed. This suggests that a better strategy for improving the performance of heralded single-photon sources is to adopt an appropriate measurement scheme for the signal photons, rather than tailoring the properties of the photon-pair generation medium.
\end{abstract}

\maketitle

\section{Introduction}
The performance of many single-photon based applications, such as optical quantum computing \cite{KniLafMil01} and quantum cryptography
\cite{Rev-Qua-cryp}, critically depends on the ability to create single photons in pure spatiotemporal states and at high rates. There exist two basic types of single-photon sources to fulfill this need. The first is based on anti-bunched emission from single-emitter systems, such as atoms \cite{Single-Photon-Atom77}, ions \cite{Single-Photon-Ion87}, molecules \cite{Single-Photon-Molecule}, semiconductors \cite{Single-photon-Semiconductor}, and quantum-dots \cite{Single-photon-QD}. The second is based on ``heralding,'' in which one photon (the ``signal'') is detected and used as a trigger to herald the presence of a paired photon (the ``idler''). Sources of this type have been demonstrated mainly in guided nonlinear media, such as $\chi^{(2)}$ waveguides \cite{Heralded-Single-Photon-SPDC86,Heralding-SPDC86,Single-Photon-PDC-99,Single-Photon-PDC01,Single-Photon-PDC07,Herald-Single-Photon-08} and $\chi^{(3)}$ fibers \cite{Single-Photon-PCF05,Single-Photon-PCF08,Single-Photon-Fiber09,Single-Photon-fiber09-2}. Heralded sources using guided media have one key advantage that the photons are generated in a single well-defined spatial mode, as opposed to a mixture of modes in the case of single-emitter sources. Single photons produced in mono-mode fibers via heralding, in particular, can be losslessly coupled into standard telecom fibers \cite{CheLeeLia06}. They thus have the potential to be a valuable resource for networked quantum information processing capable of harnessing the existing fiber-based telecommunications infrastructure.

Although heralded single-photon sources have several advantages, it is challenging to achieve simultaneously a high single-photon purity (i.e., high heralding efficiency) and a high production rate, as such sources are constrained by a fundamental trade-off between the two. For example, a common method for achieving high purity is to apply narrowband spectral filtering to the signal photons before they are detected \cite{ZeiHorWei97,FioVosSha02,CheLeeLia06,KalBlaMmo06,FulAliBri07}. This method rejects most of the usable pairs, resulting in a low single-photon production rate. An alternate method is to use spectrally-factorable photon-pair states \cite{spectral-disentanglement01,Taloring-FWM07,Single-Photon-PTP10,Single-Photon-PDC07,Herald-Single-Photon-08,Single-Photon-Fiber09,Tailoring-FWM09,TailoringFWM10}.
In such a method, however, one has to wait for a relatively long time before a second short-duration photon can be heralded. As a result, the repetition rate of the heralded single-photon creation is restricted, resulting in a relatively low production rate.

To overcome these difficulties, we have recently proposed a new approach to heralding pure single photons which does not rely on narrowband filtering or factorable photon-pair states \cite{HuaAltKum10}. The idea is to apply appropriate spectral and temporal filtering to the signal photons, so that when detected, they collapse onto a single spectral (temporal) state. Based on a simplified model, neglecting the effect of simultaneous multiple photon-pair emission, we have demonstrated via simulation that both a high purity and a high production rate can be simultaneously achieved, regardless of the spectral correlation properties of the paired photons.

In this paper, we extend our previous study \cite{HuaAltKum10} to include the effect of multiple photon-pair emission. Our goal is to identify the appropriate spectral and temporal filters for the various photon-pair correlations, for which the trade-off between high heralding efficiency and high production rate is optimally mitigated. To this end, we develop a multimode theory which describes the heralded creation of single photons in the presence of multipair emission. We then use the theory to numerically optimize the spectral and temporal filter widths for a variety of photon-pair correlations. The results reveal that while the optimization conditions vary for different correlations, the final performance using optimized filters is always quite similar. This interesting fact suggests that the key to achieving high performance in the heralded creation of single photons is to apply an appropriate measurement scheme to the signal photons. This approach is distinctly superior to the much recent effort that has been devoted to tailoring the phase-matching properties of nonlinear media in order to obtain spectrally factorable photon-pair states \cite{spectral-disentanglement01,Taloring-FWM07,Single-Photon-PTP10,Single-Photon-PDC07,Herald-Single-Photon-08,Single-Photon-Fiber09,Tailoring-FWM09,TailoringFWM10}.

The paper is organized as follows: we present the multimode heralding theory in Sec.~\ref{model}, study the optimization of measurement schemes in Sec.~\ref{odt}, and then conclude briefly in Sec.~\ref{ccls}.

\section{Multimode Theory of Heralding Single Photons}
\label{model}

In this section we develop a multimode theory modeling the heralded creation of single photons that generally applies to $\chi^{(2)}$ waveguides and $\chi^{(3)}$ fibers \cite{Raman}. In this model, photon pairs are created via spontaneous parametric down conversion (SPDC) or spontaneous four-wave mixing (SFWM) driven by pulsed pumps. The signal photons are passed through spectral and temporal filters before being measured by an on-off detector \cite{detector}. A ``click'' of the detector heralds the presence of at least one idler photon. For the sake of computational simplicity, we assume a rectangular-shaped spectral filter with an angular-frequency bandwidth $2\pi B$, where $B$ is in Hz. For the temporal filter, we consider a rectangular shutter with a $T$-second time window that is shorter than the inherent time resolution of the detector. Such a time shutter can be built using, for example, an ultrafast all-optical quantum switch \cite{HalAltKum10,HalAltKum10-2}.

We first compute the output photon-pair states from a waveguide or a fiber pumped by ligh pulses. Linearized around perfect phase-matching frequencies \cite{Taloring-FWM07}, the effective Hamiltonian describing photon-pair generation via SPDC or SFWM is given in the undepleted-pump approximation by \cite{ManWol95,FWM-Raman07}
\begin{eqnarray}
\label{eq1}
    \hat{H}(z)&=& \hbar\kappa \int d\nu_s~ d\nu_i~\phi(\nu_s+\nu_i) e^{i(\beta_s\nu_s+\beta_i\nu_i)z} \nl
    & & \times\ha_s^\dag(\nu_s)\ha_i^\dag(\nu_i)+\mathrm{H.c.}
\end{eqnarray}
Here, $\kappa$ is proportional to the pair-generation efficiency; $2\pi\nu_s$ and $2\pi\nu_i$ are the angular-frequency detunings of the signal and idler photons, respectively, from the phase-matching frequencies; and the sum-frequency profile $\phi(\nu_s+\nu_i)$ is determined by the pump spectrum. For this paper we consider
\begin{equation}
    \phi(\nu_s+\nu_i)=e^{-(\nu_s+\nu_i)^2/2\sigma^2},
\end{equation}
where $2\sigma$ is the profile bandwidth in Hz, which equals the full width at $e^{-1}$ ($e^{-2}$) maximum of the pump spectrum in the case of SPDC (SFWM). In Eq.~(\ref{eq1}), $\beta_{s,i}$ are related to the group-velocity dispersion of the signal and idler photons, respectively; and $\hat{a}^\dag_s (\nu_s)$ and $\hat{a}_i^\dag(\nu_i)$ are the creation operators, respectively, for the signal and idler photons, satisfying $[\hat{a}_p(\nu_p),\hat{a}_q^\dag(\nu'_q)]=\delta_{p,q}\delta(\nu_p-\nu'_q)$ for $p,q=s,i$. With this Hamiltonian, the quantum state of the output photon pairs from a source of length $L$ can be obtained perturbatively via
\begin{eqnarray}
    |\Psi\rangle &=&|\mathrm{vac}\rangle-\frac{i}{\hbar}\int^{L/2}_{-L/2} dz~\hat{H}(z) |\mathrm{vac}\rangle\\
    & & -\frac{1}{\hbar^2}\int^{L/2}_{-L/2} dz\int^{z}_{-L/2} dz'~\hat{H}(z)\hat{H}(z') |\mathrm{vac}\rangle+\ldots\nonumber
\end{eqnarray}
After some algebra, we find to the second-order in perturbation that
\begin{eqnarray}
\label{Psi1}
   & &  |\Psi\rangle=\mathcal{N}|\mathrm{vac}\rangle- i\kappa L \int d\nu_s ~d\nu_i~\Phi^{(1)}(\nu_s,\nu_i) |\nu_s,\nu_i\rangle\nl
    & &~~~~~~~~ +i (\kappa L)^2\int d\nu_s~ d\nu_i ~d\nu'_s ~d\nu'_i\nl
    & & ~~~~~~~~~~~~~~~~\times \Phi^{(2)}(\nu_s,\nu_s',\nu_i,\nu'_i) |\nu_s,\nu'_s,\nu_i,\nu'_i\rangle,
\end{eqnarray}
where the coefficient $\mathcal{N}$ is determined self-consistently  such that $|\Psi\rangle$ is normalized. The states $|\nu_s,\nu_i\rangle\equiv \ha^\dag_s(\nu_s)\ha^\dag_i(\nu_i)|\mathrm{vac}\rangle$ and $|\nu_s,\nu'_s,\nu_i,\nu'_i\rangle\equiv\ha^\dag_s(\nu_s)\ha^\dag_s(\nu'_s)
\ha^\dag_i(\nu_i)\ha^\dag_i(\nu'_i)|\mathrm{vac}\rangle$ represent the bases containing a single and double pairs of photons, respectively. $\Phi^{(1)}(\nu_s,\nu_i)$ and $\Phi^{(2)}(\nu_s,\nu_i)$ are the joint two-photon (single-pair) and four-photon (double-pair) spectral functions, respectively, defined as
\begin{eqnarray}
\label{psi2}
& &    \Phi^{(1)}(\nu_s,\nu_i)= \phi(\nu_s+\nu_i) \mathrm{sinc}\left(\mu_s \nu_s+\mu_i\nu_i\right),\\
\label{psi22}
& &   \Phi^{(2)}(\nu_s,\nu'_s,\nu_i,\nu'_i)=\frac{\phi(\nu_s+\nu_i) \phi(\nu'_s+\nu'_i)}{2(\mu_s\nu_s'+\mu_i\nu'_i)} \\
& & ~~~~~~\times\Big[\mathrm{sinc}[\mu_s (\nu_s+\nu'_s)+\mu_i(\nu_i+\nu'_i)]\nl
& &
~~~~~~~~~~-e^{-i(\mu_s\nu'_s+\mu_i\nu'_i)}\mathrm{sinc}\left(\mu_s \nu_s+\mu_i\nu_i\right)\Big], \nonumber
\end{eqnarray}
where $\mu_{s,i}=\beta_{s,i}L/2$ are phase-matching coefficients. From Eq.~(\ref{Psi1}), the probability to generate photon pairs is
\begin{equation}
\label{p}
    P=(\kappa L)^2 p+(\kappa L)^4 p^2
\end{equation}
with
\begin{eqnarray}
    p &=& \int d\nu_s~d\nu_i~|\Phi^{(1)}(\nu_s,\nu_i)|^2.
\end{eqnarray}


We next model the heralding stage. Existing studies have been based on the Schmidt-decomposition analysis, in which the joint two-photon spectral function (\ref{psi2}) is expanded onto a set of orthogonal Schmidt modes. The purity of heralded single-photon states is then estimated from the expansion coefficients \cite{spectral-disentanglement01,Taloring-FWM07,Herald-Single-Photon-08,
TailoringFWM10}. This analysis, however, is inapplicable to photon-pair states containing more than one pair of photons. Moreover, it is not physically rigorous because such Schmidt modes in general are not the eigenmodes of the measurement apparatus for the signal photons. Instead, a rigorous analysis must be developed following the quantum-measurement postulate, where a detection event collapses the signal-photon state onto an eigenstate of the measurement apparatus. For the present setup, such eigenstates correspond to a set of time- and bandwidth- limited modes that are chosen by following the standard procedure for the detection of band-limited signals over the measurement time \cite{PrSp61,ZhuCav90}. In the frequency domain, the measurement eigenstates containing one and two photons are given by \cite{SasSuz06,HuaAltKum10}
\begin{eqnarray}
\label{1m}
     |1\rangle_{m} &=& \int^{B/2}_{-B/2} d\nu_s~ \varphi_m(c,\nu_s) |\nu_s\rangle, \\
    \label{2m}
  |2\rangle_{m} &=& \frac{1}{\sqrt{2}} \int^{B/2}_{-B/2}  d\nu_s \int^{B/2}_{-B/2}  d\nu'_s \nl
    & & ~~~~~\times \varphi_m(c,\nu_s) \varphi_m(c,\nu'_s) |\nu_s,\!\nu'_s\rangle,
\end{eqnarray}
where $m=0,1,\ldots$ is the order number of the eigenstates, and $c=\pi B T/2$ is a dimensionless parameter determining the mode structure. The mode function
\begin{equation}
  \varphi_m(c,\nu_s)=\sqrt{\frac{(2m+1)}{B}} S_{0m}(c,\frac{2\nu_s}{B}),
\end{equation}
where $\{S_{nm}(x,y)\}$ are the angular prolate spheroidal functions. Its corresponding eigenvalue is
\begin{equation}
 \chi_m(c)=\frac{2 c}{\pi} \left(R^{(1)}_{0m}(c,1)\right)^2\le 1,
\end{equation}
where $R^{(1)}_{nm}(c,x)$ is the radial prolate spheroidal function. Ordering $\chi_0(c)>\chi_1(c)>\chi_2(c)\cdots$,
$m=0$ represents the fundamental detection mode of our interest.

With the eigenstates (\ref{1m}) and (\ref{2m}), the positive operator-valued measure (POVM) for registering signal photons by an on/off detector is given by
\begin{eqnarray}
\label{pon}
    & & \hat{\mathrm{P}}_\mathrm{on}=\sum_m \Big(\eta_{m} |1\rangle_{mm}\langle 1|+(2\eta_m-\eta^2_m)|2\rangle_{mm}\langle 2|\Big) \\
    & &~~~+\sum_{m<m'} (\eta_{m}+\eta_{m'}-\eta_{m} \eta_{m'}) |1\rangle_{mm}\langle 1|\otimes |1\rangle_{m'm'}\langle 1|, \nonumber
\end{eqnarray}
where $\eta_m=\eta \chi_m(c)$ and $\eta$ is the total detection efficiency including transmission losses and the inherent quantum efficiency of the detector. In arriving at this POVM, we have assumed that at most two photons reach the detector simultaneously, consistent with our previous approximation that at most two pairs of photons can be created per pump pulse. Furthermore, we have neglected the effects of both dark counts and after-pulsing in the detector.

With $\hat{\mathrm{P}}_\mathrm{on}$ in Eq.~(\ref{pon}), the probability of a detector click is computed via $\mathcal{P}_s =\mathrm{Tr}\{\hat{\mathrm{P}}_\mathrm{on} |\Psi\rangle \langle \Psi|\}_{s,i}$, where the trace is carried over both the signal and idler photon states. From Eqs.~(\ref{Psi1})--(\ref{psi22}) and (\ref{1m})--(\ref{pon}), we obtain
\begin{eqnarray}
\label{ps}
    \mathcal{P}_s=(\kappa L)^2\mathcal{P}^{(1)}_s+(\kappa L)^4\mathcal{P}^{(2)}_s,
\end{eqnarray}
where
\begin{eqnarray}
    & &\mathcal{P}^{(1)}_s = \sum_m\eta_m \int d\nu_i |\psi_m(\nu_i)|^2, \\
    & &\mathcal{P}^{(2)}_s = \int d\nu_i~ d\nu'_i\bigg(\sum_{m<m'} (\eta_{m}+\eta_{m'}-\eta_{m} \eta_{m'}) \\
    & & \times |\psi_{m,m'}(\nu_i,\nu'_i)|^2 +\sum_{m}(\eta_m-\eta_m^2/2)|\psi_{m,m}(\nu_i,\nu'_i)|^2 \bigg). \nonumber
\end{eqnarray}
Here, $\psi_{m}(\nu_i)$ and $\psi_{m,m'}(\nu_i,\nu'_i)$ are the (un-normalized) heralded one-photon and two-photon wavefunctions for the idler photons, respectively, defined as
\begin{eqnarray}
\label{phim}
  & & \psi^{(1)}_{m}(\nu_i)=\int^{B/2}_{-B/2} d\nu_s~ \varphi_{m}(c,\nu_s) \Phi^{(1)} (\nu_s,\nu_i), \\
& & \psi^{(2)}_{m,m'}(\nu_i,\nu'_i)=\int^{B/2}_{-B/2} d\nu_s \int^{B/2}_{-B/2} d\nu'_s ~\Phi^{(2)}(\nu_s,\nu_s',\nu_i,\nu'_i)\nl
    & & ~~~~~\Big(\varphi_{m}(c,\nu_s) \varphi_{m'}(c,\nu'_s)+
    \varphi_{m}(c,\nu'_s) \varphi_{m'}(c,\nu_s)\Big).
\end{eqnarray}

From Eqs.~(\ref{p}) and (\ref{ps}), the detection efficiency of signal photons, i.e., the probability for the signal-photon detector to click given that at least one photon pair is emitted, is given by
\begin{equation}
    \mathcal{D}_s=\frac{\mathcal{P}_s}{P}~.
\end{equation}
After a detector click, one or more idler photons are heralded, whose reduced density matrix is given by $\hat{\rho}_i=\mathrm{Tr}\{\hat{\mathrm{P}}_\mathrm{on} |\Psi\rangle \langle \Psi|\}_{s}/\mathcal{P}_s$, where the trace is carried over the signal-photon states only. After some algebra, we obtain
\begin{eqnarray}
\label{rhoi}
    \hat{\rho}_i &=&\frac{(\kappa L)^2\mathcal{P}_s^{(1)}}{\mathcal{P}_s} \hat{\rho}^{(1)}_i +\frac{(\kappa L)^4\mathcal{P}_s^{(2)}}{\mathcal{P}_s} \hat{\rho}^{(2)}_i,
\end{eqnarray}
where $\hat{\rho}^{(1)}_i$ and $\hat{\rho}^{(2)}_i$ are the normalized reduced density matrices describing the idler-photon states containing one and two photons, respectively. They satisfy $\mathrm{Tr}\{\hat{\rho}^{(1), (2)}\}_i=1$. For the one-photon density matrix, we explicitly obtain
\begin{equation}
    \hat{\rho}^{(1)}_i=\frac{1}{\mathcal{P}_s^{(1)}}\sum_{m} \eta_{m} \int d\nu_i ~ d\nu_i'~ \psi_m(\nu_i) \psi^\ast_m(\nu_i') ~|\nu_i\rangle\langle\nu_i'|.
\end{equation}

In order to characterize the properties of the heralded photons, we expand $\hat{\rho}^{(1)}_i$ onto a set of eigenstates $\{|n\rangle_i\}$,  where $\hat{\rho}^{(1)}_i=\sum^\infty_{n=0} \lambda_n |n\rangle_{ii}\langle n|$, with ordered eigenvalues $\{\lambda_i\}$ such that $\lambda_0>\lambda_1>\lambda_2\cdots $. We then compute the heralding efficiency $\mathcal{H}$ defined as the probability of finding the idler photon in the single-photon state $|0\rangle_i$ upon a detector click. From Eq.~(\ref{rhoi}), we obtain
\begin{equation}
\label{H}
    \mathcal{H}=\frac{\mathcal{P}_s^{(1)}}
    {\mathcal{P}_s^{(1)}+(\kappa L)^2 \mathcal{P}_s^{(2)}}\lambda_0.
\end{equation}
We note that there is an alternate definition of heralding efficiency, which is the probability that an idler photon can be detected after heralding. That definition, however, does not capture the quality of the heralded photons and some measurement giving the purity of the heralded state must be specified. In contrast, the metric $\mathcal{H}$ defined in Eq.~(\ref{H}) captures both the detection efficiency and the purity of the heralded photons.

The other important characteristic is the maximally-achievable production rate $\mathcal{R}$ of the heralded photons, which can be defined as
\begin{eqnarray}
\label{R}
    \mathcal{R}=\frac{(\kappa L)^2 \mathcal{P}_s^{(1)}+(\kappa L)^4 \mathcal{P}_s^{(2)}}{T_\mathrm{min}},
\end{eqnarray}
where $T_\mathrm{min}$ is the minimum amount of time required for a single heralding cycle. In practice, it is given by the largest of the following: the detection window $T$, the temporal length of the pump pulses, the temporal length of the signal photons after the filter, and the underlying pulse length of the heralded state $|0\rangle_i$. For practical considerations, in this paper we define explicitly
\begin{equation}
    T_\mathrm{min}=\max[T,4\tau_p, 4\tau_s, 4\tau_0],
\end{equation}
where $\tau_{p}$, $\tau_{s}$, and $\tau_{0}$ are the coherence times of the pump pulses, the filtered signal photons, and the heralded state $|0\rangle_i$, respectively (see Ref.~\cite{ManWol95} for an explicit definition of the coherence time).

From Eqs.~(\ref{H}) and (\ref{R}), it is clear that there is an inherent trade-off between the heralding efficiency $\mathcal{H}$ and the single-photon production rate $\mathcal{R}$. Indeed, for given pump bandwidth $\sigma$, phase-matching coefficients $\mu_{s,i}$, filtering bandwidth $B$, and measurement window $T$, the values of  $\mathcal{P}_s^{(1)}$, $\mathcal{P}_s^{(2)}$, and $T_\mathrm{min}$ are fixed. Then, $\mathcal{R}$ increases monotonically with $\kappa L$. In contrast, $\mathcal{H}$ decreases monotonically with $\kappa L$, due to the growing background emission of multiple photon pairs. Hence, one must sacrifice the production rate $\mathcal{R}$ for a higher heralding efficiency $\mathcal{H}$, or vice-verse.

Lastly, in the weak pumping regime where almost all single-photon sources operate in, the probability of emitting multiple photon pairs is much smaller than that of emitting a single pair. Thus to a good approximation,
\begin{eqnarray}
    \mathcal{H}&=&\left(1-(\kappa L)^2 \frac{\mathcal{P}^{(2)}_s}{\mathcal{P}^{(1)}_s}\right)\lambda_0, \\
    \mathcal{R}&=&(\kappa L)^2 \frac{ \mathcal{P}_s^{(1)}}{T_\mathrm{min}}.
\end{eqnarray}
Hence, the trade-off between $\mathcal{H}$ and $\mathcal{R}$ is approximately linear.

\section{Optimization of Heralding}
\label{odt}

\begin{figure}
\centering \epsfig{figure=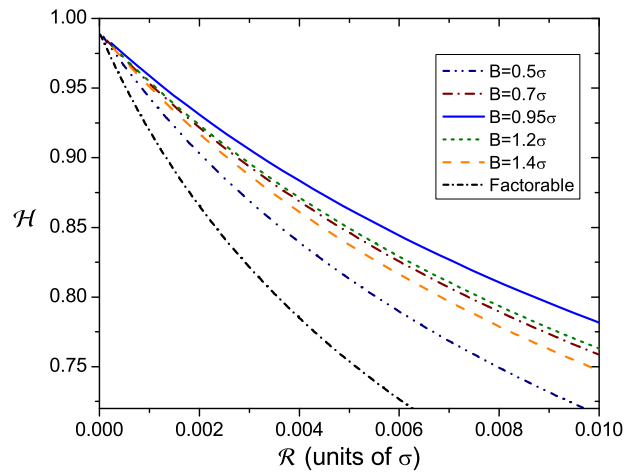, width=8.2cm} \caption{(Color online) $\mathcal{H}$ as a function of $\mathcal{R}$ for various $B$, shown for a non-factorable state with $\mu_s=\mu_i=0$. The bottom line represents the result when a highly-factorable state with $\mu_s=25/\sigma$ and $\mu_i=0$ is used but neither spectral nor temporal filtering is applied. \label{fig2}}
\end{figure}

Using the multimode theory developed in the previous section, we now study the optimization of measurement schemes for the best heralding performance in the presence of various photon-pair spectral correlations. This involves numerically identifying the appropriate spectral and temporal filtering windows which simultaneously maximize the heralding efficiency $\mathcal{H}$ and the production rate $\mathcal{R}$.

We first consider the photon-pair state (\ref{Psi1}) with $\mu_s=\mu_i=0$. This state exists approximately in a variety of photon-pair sources where the bandwidths of the phase-matching spectra are much broader than those of the pump and the signal spectral filter \cite{HalAltKum09,MedAltHal10}. The two-photon (single-pair) state in this case is not factorable according to the Schmidt decomposition analysis, with the degree-of-factorization $K\rightarrow 0$ in the limit of large $B$ \cite{Kdefine08}. In Fig.~\ref{fig2} we plot the resulting heralding efficiency $\mathcal{H}$ versus the production rate $\mathcal{R}$ for various $B$. For the sake of comparison, for each $B$ we choose $T$ to maximize $\mathcal{R}$ while fulfilling $\mathcal{H}\ge 0.99$ in the weak-pump limit. Such chosen $T$'s in fact lead to an overall optimal trade-off behavior for $0.9 \lesssim \mathcal{H}\lesssim 0.99$, as we will show later in this section. As shown in Fig.~\ref{fig2}, for each $B$, $\mathcal{H}$ decreases monotonically with $\mathcal{R}$, exhibiting the trade-off effect. The optimal trade-off, represented by the topmost $\mathcal{H}$-$\mathcal{R}$ curve in the figure, is achieved with $B=0.95\sigma$. For larger or smaller $B$'s, the $\mathcal{H}$-$\mathcal{R}$ curves fall below the optimum, showing reduced heralding performance.

For comparison, in Fig.~\ref{fig2} we have also plotted the result obtained for a factorable two-photon state without applying any spectral or temporal filtering \cite{spectral-disentanglement01,Taloring-FWM07,Herald-Single-Photon-08, TailoringFWM10}.
We have chosen $\mu_s=25/\sigma$ and $\mu_i=0$ such that in the weak-pump limit the heralding efficiency $\mathcal{H}=0.99$. These phase-matching parameters correspond to the signal photon traveling at a much slower group velocity than the pump pulse, while the idler photon travels at the pump-pulse velocity. In the joint two-photon spectrum, the underlying two-photon state corresponds to an ellipse squeezed along the axis of the signal-photon frequency \cite{Tailoring-FWM09}.
As shown, the heralding performance in this case is lower than is achievable with the non-factorable state by applying spectral and temporary filtering. \emph{This result suggests that the use of factorable states alone does not lead to an optimized trade-off between $\mathcal{H}$ and $\mathcal{R}$.}

\begin{figure}
\centering \epsfig{figure=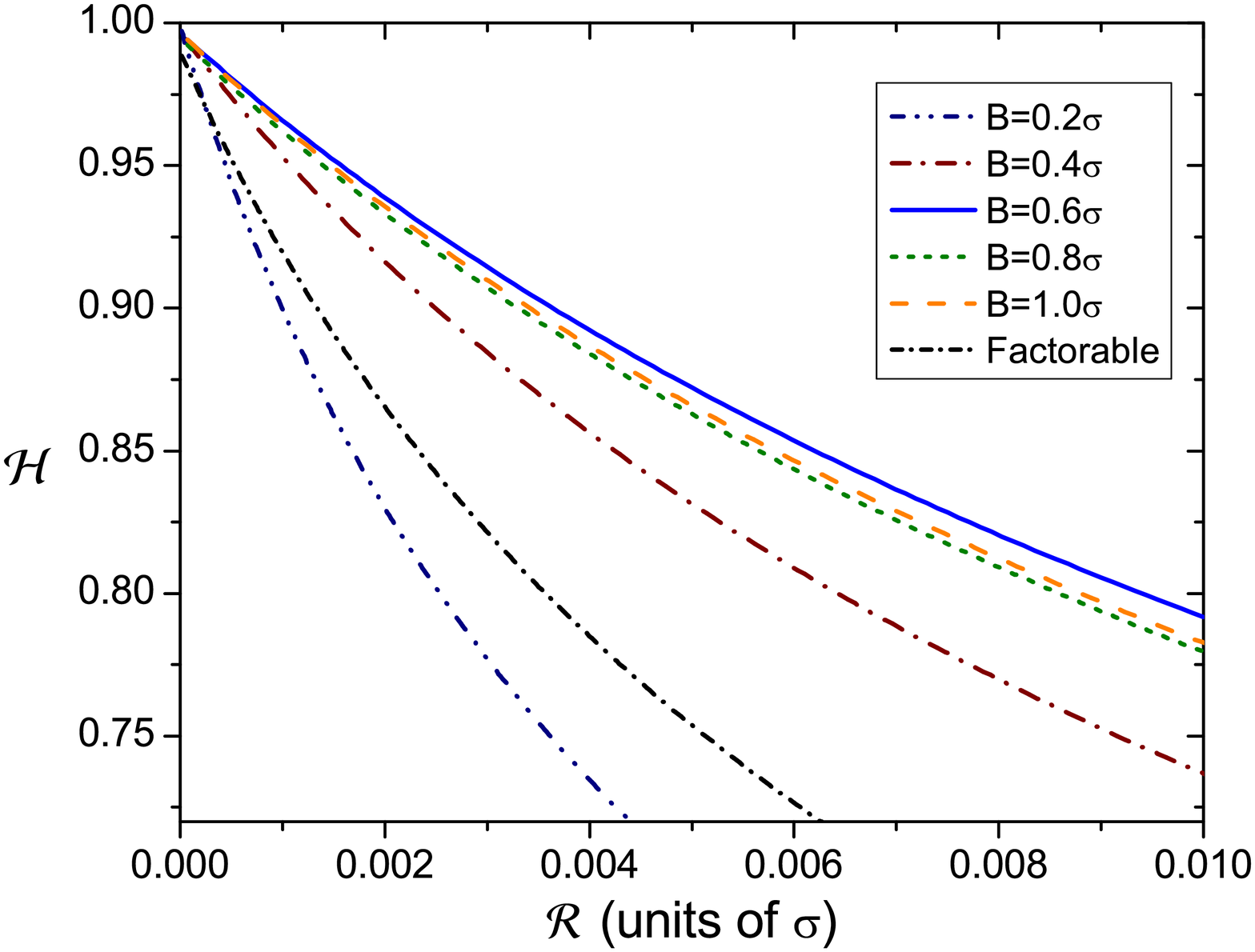, width=8.2cm} \caption{(Color online) Similar to Fig.~\ref{fig2}, but shown for a factorable photon-pair state with $\mu_s=10/\sigma$ and $\mu_i=0$. \label{fig3}}
\end{figure}

For the second example, we analyze the heralding performance using the photon-pair state (\ref{Psi1}) with $\mu_s=10/\sigma$ and $\mu_i=0$. The two-photon state in this case, similar to that described in the previous paragraph, corresponds to an ellipse in the joint two-photon spectrum. It is factorable according to the Schmidt decomposition, with the degree-of-factorability $K=1.03$. In Fig.~\ref{fig3}, we plot $\mathcal{H}$ versus $\mathcal{R}$ for different $B$'s, where, again, for each $B$ we optimize $T$ to maximize $\mathcal{R}$ while fulfilling $\mathcal{H}\ge 0.99$ in the weak-pump limit. As shown, a qualitatively similar trade-off behavior is found between $\mathcal{H}$ and $\mathcal{R}$ as in Fig.~\ref{fig2}. The optimal heralding performance, represented by the topmost $\mathcal{H}$-$\mathcal{R}$ curve, is achieved with $B=0.6\sigma$. For smaller $B$'s, the trade-off degrades rapidly. Note that for $B=0.2\sigma$, the $\mathcal{H}$-$\mathcal{R}$ curve falls below the reference curve obtained for a highly-factorable state with $\mu_s=25/\sigma$, $\mu_i=0$, and no filtering. This suggests that for factorable states, too much spectral filtering lowers the heralding performance. For larger $B$'s, on the other hand, the trade-off behavior degrades only slowly. Note, however, if no filtering is applied, the heralding performance will be quite low, as discussed in the previous paragraph.

\begin{figure}
\centering \epsfig{figure=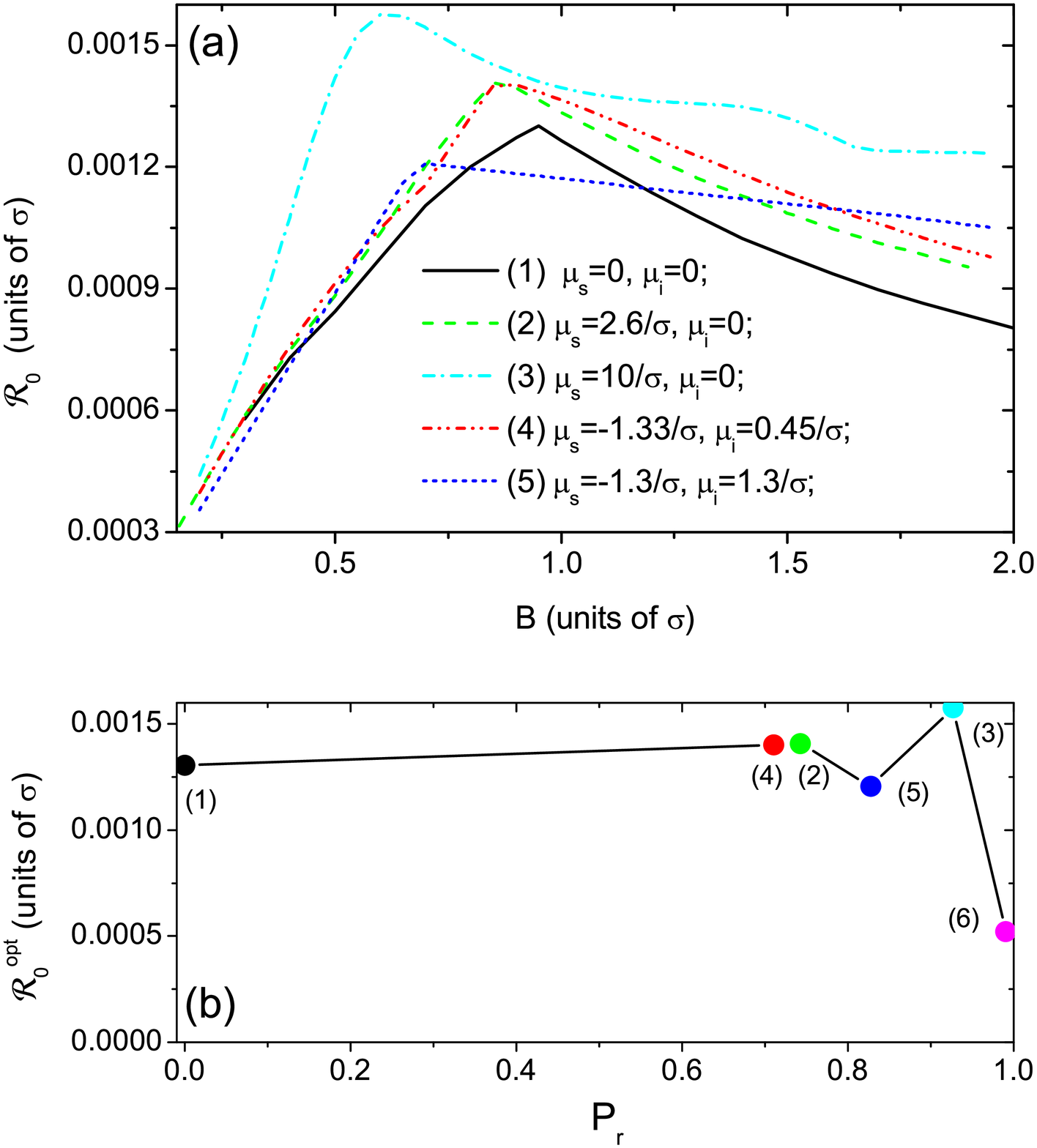, width=8.2cm} \caption{(Color online) (a) $\mathcal{R}_0$ as a function of $B$. (b) $\mathcal{R}^\mathrm{opt}_0$ versus the photon-pair purity $P_r$.  \label{fig4}}
\end{figure}

Comparing Figs.~\ref{fig2} and \ref{fig3}, we see that for a similar filter-bandwidth $B$, the trade-off behavior between $\mathcal{H}$ and $\mathcal{R}$ is different for different photon-pair correlations. By applying the optimized measurement for each correlation, however, the best trade-off behavior that can be achieved turns out to be quite similar, as evidenced by the similarity between the topmost curves in Figs.~\ref{fig2} and \ref{fig3}. This result suggests that there is no inherent relation between the achievable heralding performance and the spectral-correlation properties of the photon pairs used for heralding. To show this clearly, in Fig.~\ref{fig4}(a) we plot the production rate $\mathcal{R}_0$ as a function of $B$ for different pair correlations, with the pump power chosen such that the heralding efficiency $\mathcal{H}=0.95$ in each case. As shown, for $\mu_s=\mu_i=0$, $\mathcal{R}_0$ is peaked at $B=0.95\sigma$, for which the optimal rate $\mathcal{R}^\mathrm{opt}_0=0.0013\sigma$ is achieved. To obtain this optimum, the time window is set at $T=1.1/\sigma$. The single-pair generation rate is 5.7\% per pump pulse (the double-pair rate is 0.32\%) and the heralding repetition rate is $1/T_\mathrm{min}=\sigma/2.7$.  The detection probability $\mathcal{D}_s$, assuming a total detection efficiency of $\eta=10\%$, is 5.9\%. For $B< 0.95\sigma$, the system enters the narrowband spectral filtering regime where $\mathcal{R}_0$ decreases rapidly with $B$. In the opposite broadband filtering regime where $B> 0.95\sigma$, $T$ must be quite small in order to achieve a high heralding efficiency \cite{HuaAltKum10}. In effect, the system is then operated in the tight temporal-filtering regime \cite{HalBevGis07} with $\mathcal{R}_0$ decreasing for increasing $B$. Hence, for non-factorable states, very tight spectral or temporal filtering will lead to poor trade-off behavior between the heralding efficiency and the production rate. Only for an appropriate combination of moderate spectral and temporal filtering can both high heralding efficiency and high production rate be simultaneously achieved.

In Fig.~\ref{fig4}(a), for the factorable state with $\mu_s=10/\sigma$ and $\mu_i=0$, the optimal production rate  $\mathcal{R}^\mathrm{opt}_0=0.0016\sigma$ is achieved with $B=0.6 \sigma$. At this optimum, the single-pair emission rate is 8.9\% per pump pulse (the double-pair rate is 0.8\%) and the heralding repetition rate is $\sigma/3.2$. The detection window $T=3.2/\sigma$ and the detection efficiency $\mathcal{D}_s=5.3\%$ for $\eta=10\%$. For $B<0.6 \sigma$, $\mathcal{R}_0$ decreases rapidly with $B$. On the other hand, for $B>0.6\sigma$, $\mathcal{R}_0$ decreases relatively slowly when $B$ increases. This behavior shows that a precise control of spectral filtering is not required for factorable photon-pair states. In comparison, for a less factorable state with $\mu_s=2.6/\sigma$ and $\mu_i=0$, a similar behavior is shown in Fig.~\ref{fig4}(a) but with $\mathcal{R}^\mathrm{opt}_0=0.0014\sigma$ that is achieved at $B=0.85\sigma$.

Also in Fig.~\ref{fig4}(a), we have plotted the results for another type of factorable photon-pair states, which, instead of ellipses, correspond to circles in the joint two-photon spectrum \cite{spectral-disentanglement01}. We consider two states of this type. The first is with $\mu_s=-1.33/\sigma$ and $\mu_i=0.45/\sigma$, which could be created in potassium-titanyl-phosphate waveguides \cite{Single-Photon-PTP10}. As shown, the optimal $\mathcal{R}^\mathrm{opt}_0=0.0014\sigma$, which is achieved with $B=0.85\sigma$. The second state corresponds to $\mu_s=-1.3/\sigma$ and $\mu_i=1.3/\sigma$, which could be generated in photonic-crystal fibers \cite{Taloring-FWM07}. As shown, the optimal $\mathcal{R}^\mathrm{opt}_0=0.0012\sigma$, achieved with $B=0.7\sigma$.

As seen in Fig.~\ref{fig4}(a), while the optimization conditions vary significantly with the type of spectral correlations present in the photon pairs, the optimized single-photon production rates are more or less the same regardless of the correlation properties. This result suggests that by adopting optimized filtering, any spectral correlation can lead to a similarly high heralding performance. In other words, increasing the factorability of the underlying photon-pair states would not (significantly) improve the heralding performance. To clearly show this, in Fig.~\ref{fig4}(b) we plot $\mathcal{R}^\mathrm{opt}_0$ versus the purity $P_r=1/K$ of the two-photon (single-pair) states calculated via the Schmidt decomposition \cite{Kdefine08}. Each of the data points (1)-(5) correspond to the photon-pair states (1)-(5) listed in Fig.~\ref{fig4}(a). Point (6) is the result for the highly-factorable photon-pair state with $\mu_s=25/\sigma$ and $\mu_i=0$, without applying any spectral or temporal filtering. As shown, despite a large variance in the factorability of the photon-pair states, ranging from $P_r=0$ to $P_r\approx 1$, the maximum production rates remain nearly the same when the optimized measurement is employed for each correlation. For a highly-factorable state but without filtering, in contrast,  $\mathcal{R}^\mathrm{opt}_0$ is much smaller, as shown by the data point (6) in Fig.~\ref{fig4}(b).

Thus far, we have studied the optimization of measurements for heralding by surveying $B$. For each $B$, we have chosen $T$ to maximize $\mathcal{R}$ while maintaining $\mathcal{H}\ge 0.99$ in the weak-pump limit. Such $B$ and $T$, strictly speaking, may not give the maximum $\mathcal{R}$ for every $\mathcal{H}$, or vice-versa. In practice, however, the figure-of-merit for a single-photon source is the relation between $\mathcal{R}$ and $\mathcal{H}$, which ultimately limits the performance of a single-photon-based application. A well-known example is the BB84 quantum cryptography, where the obtainable fresh-key generation rate is limited fundamentally by the trade-off between the quantum-bit-error rate, which is determined by $\mathcal{H}$, and the raw-key rate, which is determined by $\mathcal{R}$ \cite{GisRibTit02}. A goal of this paper is thus to study how to optimally mitigate the $\mathcal{R}$-$\mathcal{H}$ trade-off in order to achieve the best performance in such applications.

\begin{figure}
\centering \epsfig{figure=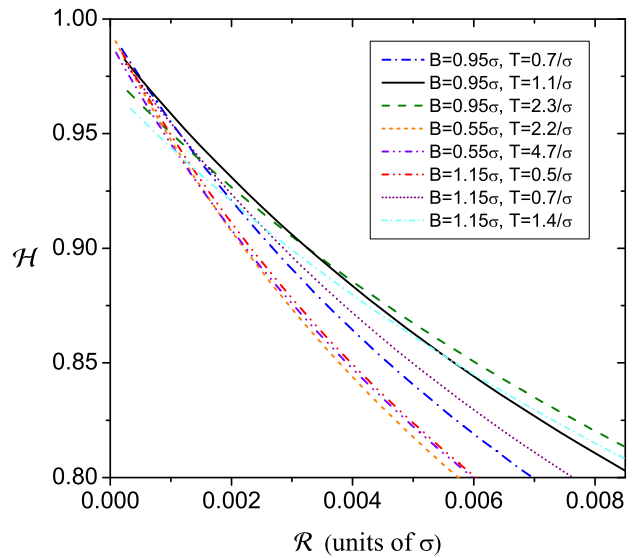, width=8.2cm} \caption{(Color online) The heralding efficiency $\mathcal{H}$ as a function of the single-photon production rate $\mathcal{R}$ for $\mu_s=\mu_i=0$ and a variety of $B$'s and $T$'s. \label{fig5}}
\end{figure}
To show that the optimal trade-off behavior is indeed achieved, in Fig.~\ref{fig5} we plot $\mathcal{H}$ versus $\mathcal{R}$ for various choices of $B$ and $T$, with $\mu_s=\mu_i=0$ as considered in Fig.~\ref{fig2}. For $B=0.95\sigma$ and $T=0.7/\sigma$, $\mathcal{H}$ approaches $0.995$ in the limit of small $\mathcal{R}$. However, it decays rapidly as $\mathcal{R}$ increases. Hence, a small $T$ is superior for pair generation at very-high heralding efficiencies but low production rates. For $B=0.95\sigma$ and $T=2.3/\sigma$, in contrast, $\mathcal{H}$ is $0.97$ in the small $\mathcal{R}$ limit. Yet, it decays more slowly with $\mathcal{R}$. A large $T$ is thus advantageous for pair generation at relatively lower heralding efficiencies but higher production rates. A moderate $T$, therefore, will give rise to a more balanced trade-off behavior in the two regimes. For $T=1.1/\sigma$, in particular, the $\mathcal{H}$-$\mathcal{R}$ curve is close to optimal in both regimes, exhibiting good overall heralding performance. In Fig.~\ref{fig5} we also plot the results for a variety of $B$'s and $T$'s. Comparing all the plotted curves confirms that our choice of $B=0.95\sigma$ and $T=1.1/\sigma$ indeed globally optimizes the heralding performance for $0.9\lesssim \mathcal{H}<0.99$. Lastly, we have similarly validated our optimization method for the other types of photon-pair correlations considered above.

\section{Conclusion}
\label{ccls}
We have developed a multimode theory for the description of heralded generation of single photons using waveguide or fiber-based photon-pair sources. Our theory takes into account the background emission of multiple photon pairs during each pump pulse, as well as the multimode nature of the time-bandwidth-limited measurements of photonic signals. Based on this theory, we have numerically identified the optimized measurement schemes that give rise to the best heralding performance in the presence of various photon-pair correlations. Interestingly, we have discovered that with the optimized measurement, similar heralding performance can be achieved irrespective of the spectral-correlation property of the used photon-pair source. This suggests that instead of tailoring the photon-pair sources for prescribed correlation properties, the key to improving the heralded generation of single photons is to appropriately measure, via spectral and temporal filtering, the signal photons that are detected for heralding.

This research was supported in part by the Defense
Advanced Research Projects Agency under
Grant No. W31P4Q-09-1-0014 and by the United States Air
Force Office of Scientific Research under Grant
No. FA9550-09-1-0593.


\begin{thebibliography}{42}
\bibitem[{Ram()}]{Raman}
\bibinfo{note}{The present model does not include the effect of Raman
  scattering, which is usually not negligible for existing fiber-based
  photon-pair sources. Such an effect can nevertheless be mitigated in various
  ways, including lowering the fiber temperature.}
\bibitem[{det()}]{detector}
\bibinfo{note}{Instead of an on/off detector, one can use a photon-number
  resolving detector, by which the effect of multiple photon-pair emission can
  be suppressed provided with a near-unity total detection efficiency.}
\end{thebibliography}
\end{document}